# ارائه یک مدل پیش‌بینی یال مبتنی بر شباهت ساختاری و هوموفیلی در شبکه‌های اجتماعی

علیرضا اسحاق‌پور[1]، مصطفی صالحی[1*]، وحید رنجبر[2]

[1] گروه بین‌رشته‌ای فناوری، دانشکده علوم و فنون نوین، دانشگاه تهران، تهران

[2] دانشکده مهندسی کامپیوتر، دانشگاه یزد، یزد

## چکیده

در سال‌های اخیر شبکه‌های اجتماعی مجازی روز به روز در حال رشد و تغییر می‌باشند. یال‌های جدید نشان‌دهنده‌ی تعاملات میان گره‌ها می‌باشند و پیش‌بینی آن‌ها از اهمیت بالایی برخوردار است. معیارهای پیش‌بینی یال را می‌توان به دو گروه مبتنی بر همسایگی گره و مبتنی بر پیمایش مسیر تقسیم کرد. محققان ایجاد یال جدید در شبکه را از منظر نظری به دو علت نزدیکی در گراف و هوموفیلی نسبت می‌دهند. با وجود مطالعات بسیار در حوزه علوم شبکه مطالعه تاثیر دو رویکرد نظری در کنار یکدیگر در ایجاد یال‌ها مسئله‌ای باز محسوب می‌شود و تاکنون معیارهای شباهت مبتنی بر همسایگی گره از این منظر مطالعه نشده‌اند. در این پژوهش مدلی ارائه کردیم تا با استفاده از مزایای هر دو رویکرد نزدیکی در گراف و هوموفیلی استفاده نماییم و با استفاده از آن توانستیم بر دقت معیارهای شباهت مبتنی بر همسایگی گره بیفزاییم. برای ارزیابی این پژوهش از دو مجموعه داده شبکه اجتماعی مجازی زنجان و شبکه اجتماعی مجازی پوکک استفاده شده است که مجموعه داده اول برای این پژوهش جمع‌آوری و سپس تکمیل گشته است.

### واژگان کلیدی

پیش‌بینی یال، شباهت هوموفیلی، شباهت ساختاری، شبکه‌های اجتماعی

* نویسنده عهده‌دار مکاتبات، تهران، خیابان کارگر شمالی، دانشکده علوم و فنون نوین دانشگاه تهران، طبقه سوم، اتاق ۳۳۷، تلفن تماس: ۰۲۱۸۶۰۹۳۲۹۸، نمابر: ۰۲۱۸۸۴۹۷۳۲۴، ایمیل: mostafa_salehi@ut.ac.ir

# Providing a Link Prediction Model based on Structural and Homophily Similarity in Social Networks

**Alireza Eshaghpour[1], Mostafa Salehi[1], Vahid Ranjbar[2]**

[1] Faculty of New Sciences and Technologies, Tehran University, Tehran
[2] Department of Computer Engineering, Yazd University, Yazd

**Abstract**

In recent years, with the growing number of online social networks, these networks have become one of the best markets for advertising and commerce, so studying these networks is very important. Most online social networks are growing and changing with new communications (new edges). Forecasting new edges in online social networks can give us a better understanding of the growth of these networks. Link prediction has many important applications. These include predicting future social networking interactions, the ability to manage and design useful organizational communications, and predicting and preventing relationships in terrorist gangs.

There have been many studies of link prediction in the field of engineering and humanities. Scientists attribute the existence of a new relationship between two individuals for two reasons: 1) Proximity to the graph (structure) 2) Similar properties of the two individuals (Homophile law). Based on the two approaches mentioned, many studies have been carried out and the researchers have presented different similarity metrics for each category. However, studying the impact of the two approaches working together to create new edges remains an open problem.

Similarity metrics can also be divided into two categories; Neighborhood-based and path-based. Neighborhood-based metrics have the advantage that they do not need to access the whole graph to compute, whereas the whole graph must be available at the same time to calculate path-based metrics.

So far, the above two theoretical approaches (proximity and homophile) have not been found together in the neighborhood-based metrics. In this paper, we first attempt to provide a solution to determine importance of the proximity to the graph and similar features in the connectivity of the graphs. Then obtained weights are assigned to both proximity and homophile. Then the best similarity metric in each approach are obtained. Finally, the selected metric of homophily similarity and structural similarity are combined with the obtained weights.

The results of this study were evaluated on two datasets; Zanjan University Graduate School of Social Sciences and Pokec online Social Network. The first data set was collected for this study and then the questionnaires and data collection methods were filled out. Since this dataset is one of the few Iranian datasets that has been compiled with its users' specifications, it can be of great value. In this paper, we have been able to increase the accuracy of Neighborhood-based similarity metric by using two proximity in graph and homophily approaches.

## 1- مقدمه

در دهه دوم از قرن ۲۱، شاهدیم که مجموعه‌ای از افراد از طریق کاربردهای بستر اینترنت با یکدیگر ارتباطاتی از قبیل دوستی، ارتباط اجتماعی و همکاری برقرار می‌کنند و کنش‌های اجتماعی خود را با الهام از امکانات فنی شبکه و زمینه‌های اجتماعی و سازوکارهای ارتباطاتی سامان می‌بخشند. این مجموعه از افراد و کنش‌هایی که بین آن‌ها رخ می‌دهد، شبکه‌های اجتماعی مجازی را تشکیل می‌دهند. هرچند آمار دقیقی برای تعداد کاربران شبکه‌های اجتماعی مجازی وجود ندارد، ولی با توجه به آمار تجاری، شبکه‌های اجتماعی مجازی در سال‌های اخیر توانسته‌اند تعداد کاربران بسیار بالایی را به خود جذب کنند و همین امر باعث شده است این شبکه‌ها به یکی از بهترین بازارها برای تبلیغات و تجارت تبدیل شوند [۱-۲]، بنابراین مطالعه این شبکه‌ها از اهمیت بالایی برخوردار می‌باشد [۴] اکثر این شبکه‌ها، ساختار و ویژگی‌های مشخصی دارند [۵-۶]. یکی از موضوعات مورد مطالعه در علوم شبکه طراحی مدل‌هایی است که ظهور چنین ساختارهایی را پیش‌بینی کند [۳، ۷، ۸].

بسیاری از شبکه‌های اجتماعی مجازی به شدت پویا بوده و با اضافه شدن یال‌های جدید در حال رشد و تغییر می‌باشند. یال‌های جدید نشان دهنده تعاملات میان گره‌های شبکه می‌باشد. بنابراین مطالعه شبکه‌های اجتماعی در سطح یال‌ها و گره‌ها می‌تواند درک بهتری از مکانیزم رشد شبکه‌ها در اختیار ما قرار دهد که به واسطه آن می‌توان به طراحی مدلی برای پیش‌بینی رشد شبکه پرداخت. این مطلب در علوم شبکه با اصطلاح پیش‌بینی یال مطرح می‌شود و به این معناست که با داشتن ساختار شبکه در لحظه $t$ بتوان ساختار شبکه را در آینده نزدیک در زمان $t' = t + x$ پیش‌بینی کرد.

پیش‌بینی یال کاربردهای بسیار مهمی دارد. از جمله آن می‌توان به پیش‌بینی تعاملات آینده شبکه‌های اجتماعی، توانایی مدیریت و طراحی ارتباطات سازمانی سودمند و پیش‌بینی روابط در باندهای تروریستی و پیشگیری از آن اشاره کرد [۱۰و۱۱]. از

**شکل ۱: ساختار روش پیشنهادی**

Figure 1: The structure of the proposed approach

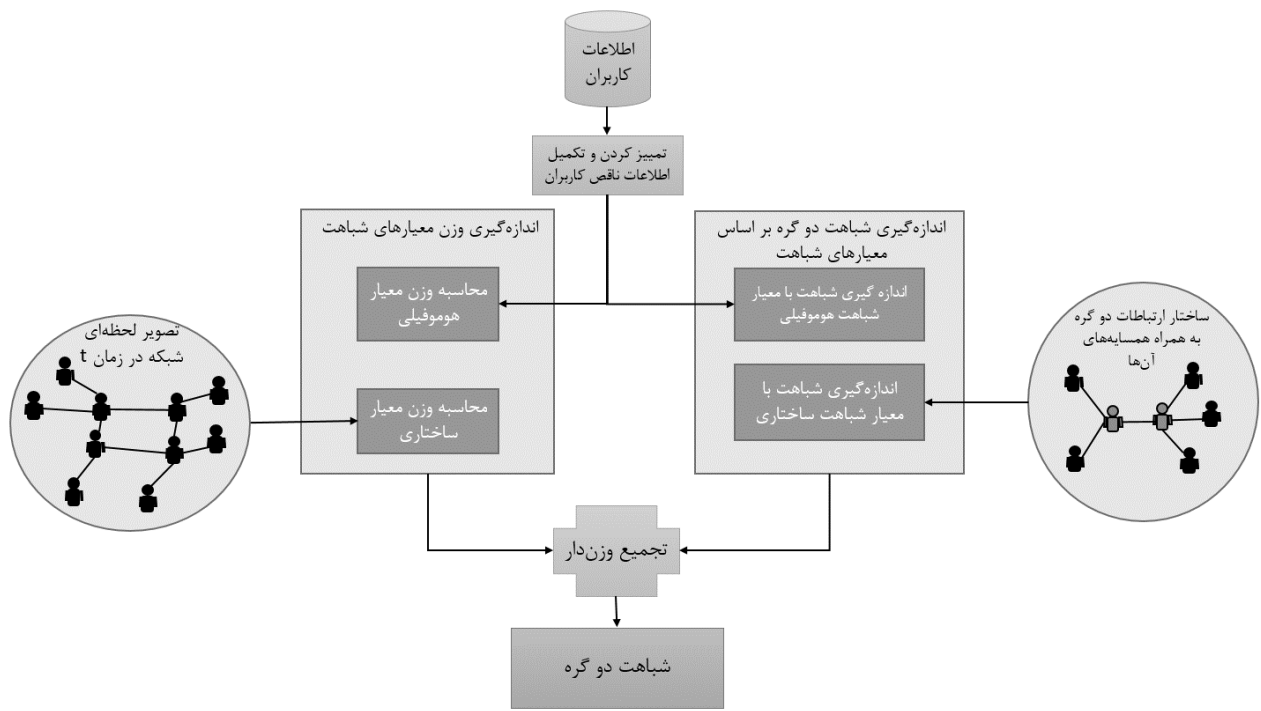

کاربردهای دیگر پیشبینی یال میتوان به استفاده از آن برای ایجاد سامانه‌های توصیه‌گر برای دوستیابی در شبکه‌های اجتماعی اشاره کرد. سامانه‌های توصیه‌گر، سامانه‌هایی هستند که به کمک اطلاعات موجود و تحلیل رفتار و خصوصیات کاربران، پیشنهاداتی خودکاری به آنها ارائه می‌دهند [۱۳].

مطالعات بسیاری در رابطه با پیش‌بینی یال در حوزه مهندسی و علوم انسانی مطرح شده است، دانشمندان وجود ارتباط جدید بین دو فرد را به دو دلیل نسبت می‌دهند: ۱- نزدیکی در گراف[1] (ساختار) ۲- ویژگی‌های مشابه دو فرد (قانون هوموفیلی[2]) [۸-۹]. بر اساس دو رویکرد فوق مطالعات بسیاری انجام شده است و محققین معیارهای شباهت متفاوتی را برای هر دسته ارائه کرده‌اند [۲۱-۲۳]. با این حال مطالعه تاثیر دو رویکرد فوق در کنار یکدیگر در ایجاد یال‌های جدید همچنان مسئله‌ای باز محسوب می‌شود.

معیارهای شباهت را می توان به دو گروه مبتنی بر همسایگی گره و مبتنی بر پیمایش مسیر تقسیم کرد. معیارهای مبتنی بر همسایگی گره این مزیت را دارند که برای محاسبه نیازی به دسترسی به کل گراف ندارند در صورتی که برای محاسبه معیارهای مبتنی بر پیمایش مسیر حتما باید کل گراف دوستی در همان لحظه ($t'$) در دسترس باشد [۱۴].

تاکنون در دسته معیارهای مبتنی بر همسایگی گره دو رویکرد نظری فوق (نزدیکی در گراف و هوموفیلی) در کنار یکدیگر دیده نشده است. در این تحقیق با ارائه یک مدل که ساختار آن در شکل ۱ ملاحظه می‌شود توانستیم با استفاده از دو رویکرد نزدیکی در گراف و هوموفیلی بر دقت معیارهای شباهت مبتنی بر

---

[1] Close in network

[2] Homophily

همســایگی گره بیفزاییم. همچنین در مدل ارائه شــده نیازی به دسترسی کل گراف در زمان t وجود نخواهد داشت و صرفا داشتن یک تصویر لحظه‌ای از شبکه در زمان گذشته (t) کفایت می‌کند.

همچنین از خروجی این پژوهش و وزن‌هایی که برای هوموفیلی شبکه و ساختار شبکه حاصل شده است می‌توان در سایر کارهای علمی نظیر انتخاب معیــار تاثیرگــذار هوموفیلی در شــبکه و تشخیص اجتماعات برخط استفاده کرد.

نتایج این پژوهش بر روی دو مجموعه داده شبکه اجتماعی مجازی دانشــگاه تحصــیلات تکمیلی زنجان و شــبکه اجتماعی مجازی پوکک ارزیابی شــده اســت. مجموعه داده اول برای این پژوهش جمع‌آوری شــده و ســپس با اســتفاده از پرسش نامه و روش‌های تکمیل اطلاعات، اطلاعات کاربری افراد تکمیل گردیده اســت. از آنجایی که این مجموعه داده جزو معدود مجموعه داده‌های ایرانی می‌باشــد که همراه با مشــخصــات کاربران آن جمع‌آوری شده است، می‌تواند از ارزش بالایی برخوردار باشد.

## ۲- پیشینه موضوع

در حوزه مهندسی و علوم انسانی پیرامون پیش‌بینی یال مطالعات زیادی انجام شــده اســت، محققان در این مطالعات وجود ارتباط جدید بین دو فرد را به دو دلیل نســبت می‌دهند: ۱- نزدیکی در گراف ۲- ویژگی‌های مشــابه دو فرد [۸-۹]. نزدیکی در گراف به این امر اشاره دارد که دو فردی که به طور مستقیم در ارتباط نیستند ولی به واسطه یک یا چند فرد (دوستان مشترک) در شبکه به صورت غیر مستقیم با یکدیگر ارتباط دارند در آینده‌ی نزدیک با احتمال بیشتری با یکدیگر ارتباط برقرار می‌کنند (به طور مثال آشــنایی آن‌ها با یکدیگر در مهمانی از طریق دوســت مشــترک). مطالعات نشان داده است که هر چقدر دو فرد در گراف دوستی به یکدیگر نزدیک‌تر باشــند در آینده نزدیک احتمال برقراری ارتباط بین آن‌ها بیشتر خواهد بود.

ویژگی‌های مشــابه دو فرد به این نکته اشــاره دارد که هر چقدر دو فرد دارای ویژگی‌های مشــابه بیشــتری باشــند، احتمال برقراری ارتباط آن دو در آینده نزدیک بیشــتر خواهد بود. از این موضــوع در ادبیات علوم انســانی با عنوان قانون هوموفیلی و در ادبیات علوم شــبکه با عنوان اختلاط همگن[۱] یاد می‌شــود. گاهی اوقات بسته به جامعه آماری، افراد تمایل دارند تا با کسانی که در یک ویژگی مشخص با آن‌ها شباهت ندارند، معاشرت کنند. به طور مثال در روابط عاشــقانه، افراد با کســانی که جنســیت متفاوتی با آن‌ها دارند معاشــرت می‌نمایند یا در جامعه خانوادگی، افراد با پدربزرگ‌ها و مادربزرگ‌ها که از نظر ســنی با آن‌ها تفاوت زیادی دارند ارتباط بیشتری برقرار می‌کنند. به این تمایل در ادبیات علوم شــبکه اختلاط ناهماهنگ[۲] و در ادبیات علوم اجتماعی هتروفیلی[۳] گفته می‌شود.

بدیهی است نزدیکی در گراف و هوموفیلی، الزاما از یکدیگر مســتقل نبوده و دارای هم‌وشــانی نیز می‌باشــند (به طور مثال بر اســاس قانون هوموفیلی دوســتان یک فرد با آن فرد ویژگی‌های

---

[1] Assortative mixing
[2] Disassortative mixing
[3] Hetrophily

مشترکی دارند بنابراین وجود دوست مشترک که مبین نزدیکی در گراف است با قانون هموفیلی نیز، قابل توجیه می‌باشد.

در حوزه مهندسی بیشتر بر روی نزدیکی در گراف تمرکز شده است و معیارهای متعددی بر این اساس ارائه شده است که تعدادی از آن‌ها به صورت خلاصه در جدول 1 ارائه شده است. (در ادامه معیارهای شباهتی که بر اساس نزدیکی در گراف محاسبه می‌شوند با عنوان معیارهای شباهت ساختاری عنوان می‌شوند). در مطالعات صورت گرفته از بین معیارهای عنوان شده معیار جاکارد و شباهت شبکه دقت بالاتری نسبت به سایر معیارها کسب کرده‌اند [25].

جدول 1: معیارهای شباهت ساختاری
Table 1: Network similarity

| معیار شباهت | سال | مرجع |
|---|---|---|
| ضریب جاکارد | 1901 | [18] |
| همبستگی نقطه به نقطه اطلاعات مشترک | 1991 | [19] |
| آدامیک و آدار | 2003 | [20] |
| کسینوس | 2004 | [24] |
| شباهت شبکه | 2013 | [12] |

در حوزه علوم انسانی محققین به مباحث نظری قوانین هموفیلی پرداخته‌اند و با ارائه شواهد و مطالعات تجربی توانسته‌اند این مبحث را غنی‌تر کنند. با این وجود مطالعات انجام شده پیرامون ارائه معیارها براساس قوانین هموفیلی نسبت به معیارهای شباهت ساختاری بسیار کمتر بوده است. تعدادی از معیارهای ارائه شده بر اساس قوانین هموفیلی در جدول 2 ارائه شده است (در ادامه به این دسته از معیارها، معیارهای شباهت هموفیلی اطلاق

می‌شود). از بین معیارهای شباهت هموفیلی معیار OF و IOF توانسته‌اند نتایج خوبی را کسب نمایند [12]. بنابراین در این پژوهش این دو معیار به عنوان معیارهای کاندید انتخاب شده‌اند.

از رویکردی دیگر می‌توان معیارهای شباهت را به دو دسته تقسیم کرد. 1- معیارهای شباهت مبتنی بر همسایگی گره 2- معیارهای شباهت مبتنی بر پیمایش مسیر. معیارهای معرفی شده در جداول 1 و 2 همگی در گروه معیارهای شباهت مبتنی بر همسایگی گره قرار داده می‌شوند. در دسته معیارهای مبتنی بر پیمایش مسیر که در سال‌های اخیر محققین به آن توجه بیشتری داشته‌اند پیش‌بینی یال بر اساس پیمودن یک مسیر (یا فرامسیر) در گراف مشخص می‌شود. بدین معنا که برای پیش‌بینی ارتباطات آتی یک گره، از آن گره الگوریتم آغاز شده و در نهایت با پیمودن مسیری در گراف به گره‌هایی از شبکه ختم می‌شود و در پایان از بین آن‌ها تعدادی گره بر اساس اولویت به عنوان کاندید ارتباطات آتی آن گره معرفی می‌شوند [26-28]. بعضی از معیارهای این دسته از ویژگی‌های شباهت در پیمودن مسیر بهره برده‌اند که به می‌توان گفت که در این معیارها از دو رویکرد نظری فوق برای پیش‌بینی یال استفاده شده است [8].

جدول 2: معیارهای شباهت هموفیلی
Table 2: Homophily somilarity

| معیار شباهت | سال | مرجع |
|---|---|---|
| همپوشانی | 1986 | [15] |
| گودال | 1966 | [16] |
| اسکین | 2002 | [17] |
| IOF | 2008 | [11] |
| OF | 2013 | [12] |

با توجه به دقت بالاتر معیارهای دسته مبتنی بر پیمایش مسیر در قیاس با معیارهای دسته مبتنی بر همسایگی گره همچنان از معیارهای مبتنی بر همسایگی گره در بسیاری از کارهای علمی استفاده می‌شود. علت آن هست که برای محاسبه معیارهای شباهت مبتنی بر همسایگی گره بر خلاف معیارهای مبتنی بر پیمایش مسیر دسترسی به کل گراف دوستی نیاز نیست و از پیچیدگی زمانی کمتری برخوردار هستند. همین امر باعث شده است که همچنان از این معیارها استفاده شود. با توجه به مزایای معیارهای شباهت مبتنی بر همسایگی گره رویکرد این پژوهش افزایش دقت این معیارها می‌باشد، لذا تمرکز اصلی این پژوهش بر روی این دسته از معیارها می‌باشد.

نکته‌ی قابل تاملی که وجود دارد این است که تاکنون در بین معیارهای مبتنی بر همسایگی گره مطالعه‌ای مبنی بر در نظر گرفتن هر دو تحلیل نظری نزدیکی در گراف و هوموفیلی در کنار یکدیگر دیده نشده است. با وجود این امر در این تحقیق ما به دنبال آن هستیم تا با ارائه یک روش علاوه بر حفظ مزیت روش‌های مبتنی بر همسایگی گره بتوانیم هوموفیلی و نزدیکی در گراف را در کنار یکدیگر در نظر بگیریم تا بتوانیم بر دقت این معیارها بیفزاییم.

## ۳- روش اندازه‌گیری

ساده‌ترین روشی که برای در نظر گرفتن هر دو رویکرد هوموفیلی و نزدیکی در گراف وجود دارد تجمیع معیارهای متناظر هر دو گروه می‌باشد. اما نکته‌ای که در اینجا مطرح می‌شود این است که با در نظر نگرفتن اهمیت هر یک از این معیارها در شبکه مورد مطالعه امکان گرفتن نتایج ضعیف‌تر وجود خواهد داشت.

در این تحقیق فرض شده است که ساختار شبکه در آینده نزدیک شامل تغییرات عمده‌ای نخواهد شد با این فرض راه حل عملی استفاده شده در این تحقیق این هست که در فاز اول ابتدا به ارزیابی شبکه مورد مطالعه پرداخته شده تا براساس آن بتوان میزان اهمیت هر یک از رویکردهای هوموفیلی و نزدیکی در گراف سنجیده شود. در واقع به دنبال آن هستیم که کاربران شبکه مورد مطالعه، تاکنون در انتخاب دوستان خود چقدر به مسئله نزدیکی در گراف و چقدر به مسئله وجود ویژگی‌های مشابه اهمیت داده‌اند. در ادامه براساس آن وزن‌هایی حاصل می‌شود. در نهایت معیارهای شباهت هوموفیلی و معیارهای شباهت ساختاری با یکدیگر با احتساب وزن‌های به دست آمده تجمیع می‌شوند. در ادامه نحوه محاسبه وزن‌ها برای هر یک از دو رویکرد هوموفیلی و نزدیکی در گراف تبیین خواهد شد.

### ۱-۳- اندازه‌گیری وزن ویژگی‌های هوموفیلی

نیومن، معیاری برای اندازه‌گیری ویژگی‌های هوموفیلی برای شبکه‌های اجتماعی مجازی، ارائه کرده است [۲۹]. از این معیار می‌توان برای مشخص کردن وزن ویژگی‌های هوموفیلی برای متغیرهای اسمی استفاده شود (در صورت وجود متغیرهای ترتیبی، فاصله‌ای و نسبی، این متغیرها باید به متغیرهای اسمی تبدیل شوند). بنابراین در این کار برای تعیین وزن ویژگی هوموفیلی از رابطه (۱) بهره گرفته شده است. این رابطه بیانگر وزن نرمال شده برای ویژگی هوموفیلی $c$ می‌باشد.

$$\frac{W(c)}{W_{max}(c)} = \frac{\sum_{ij}(A_{ij} - \frac{k_i k_j}{2m})\delta(c_i,c_j)}{2m - \sum_{ij}\frac{k_i k_j}{2m}\delta(c_i,c_j)} \quad (1)$$

$W_{mac}(c)$ بیشینه وزن محتمل برای ویژگی c را نشان می‌دهد و $W(c)$ بیانگر وزن محتمل برای ویژگی c می‌باشد. $W(c)$ از تفضیل تعداد روابطی که بر اساس ویژگی هموفیلی c در گراف دوستی حاصل شده‌اند با تعداد روابطی که بر اساس این ویژگی هموفیلی در گراف تصادفی و با توزیع درجه یکسان حاصل شده‌اند به دست می‌آید. در این رابطه $A_{i,j}$ ماتریس مجاورت گراف مورد مطالعه، $k_i$ درجه گره i و m نشان دهنده تعداد یال‌های گراف مورد مطالعه می‌باشند. همچنین $\delta(c_i,c_j)$ یک متغیر بولی می‌باشد که خروجی آن در صورت شباهت ویژگی هموفیلی دو گره i و j برابر عدد یک و در غیر اینصورت برابر عدد صفر می‌باشد. W(c) می‌تواند مثبت، منفی و صفر باشد. می‌توانیم از بازه‌های عددی وزن حاصل، نتایج زیر را بگیریم:

W(c) > 0: تعداد ارتباطاتی که در این شبکه براساس تشابه وجود دارد از تعداد ارتباطاتی که در حالت تصادفی به دست می‌آید، بیشتر می‌باشد. در واقع در شبکه مورد مطالعه تعداد بیشتری از ارتباطات براساس تشابه ویژگی c نسبت به مقدار مورد انتظار داریم. بنابراین برای ویژگی c دو نتیجه می‌توانیم بگیریم: اول آنکه در شبکه مورد مطالعه در رابطه با ویژگی c هموفیلی وجود دارد نه هتروفیلی. به طور مثال در شبکه دانشجویان افراد با تحصیلات یکسان با یکدیگر بیشتر ارتباط برقرار می‌کنند، همچنین در شبکه دوستی، افراد با گروه‌های سنی مشابه با یکدیگر ارتباط بیشتری برقرار می‌کنند. در واقع این امر بدان معناست که در این شبکه‌ها برای ویژگی تحصیلات و سن هموفیلی وجود دارد. دوم آنکه بزرگ بودن مقدار عددی وزن ویژگی c نشان دهنده‌ی تاثیرگذاری بیشتر ویژگی c در انتخاب دوستان می‌باشد.

W(c) < 0: تعداد ارتباطاتی که در این شبکه براساس تشابه وجود دارد از تعداد ارتباطاتی که در حالت تصادفی به دست می‌آید، کمتر می‌باشد. در واقع در شبکه مورد مطالعه تعداد کمتری از ارتباطات براساس تشابه ویژگی c نسبت به مقدار مورد انتظار داریم. بنابراین می‌توانیم دو نتیجه بگیریم: اول آنکه در شبکه مورد مطالعه در رابطه با ویژگی c هتروفیلی وجود دارد نه هموفیلی. دوم آنکه بزرگ بودن مقدار عددی وزن ویژگی c نشان دهنده‌ی تاثیرگذاری بیشتر ویژگی c در انتخاب دوستان می‌باشد.

W(c) = 0: از صفر بودن وزن هموفیلی برای ویژگی c می‌توان فهمید که این ویژگی تاثیری در انتخاب دوستان توسط افراد در شبکه مورد مطالعه ندارد.

به عنوان مثال (جنسیت)W برای شکل ۲ برابر ۰/۱۲ می‌باشد، این بدین معناست که در این شبکه مشابهت جنسیت باعث افزایش احتمال دوستی بین افراد می‌شود (هموفیلی) و وزن این احتمال برابر ۰/۱۲ می‌باشد.

## ۲-۳- اندازه‌گیری وزن ویژگی ساختاری

همان طور که در بخش پیشینه موضوع بیان شد الگوی دوست‌یابی افراد را می‌توان به دو قانون هموفیلی و نزدیکی در گراف، نسبت داد. در ادبیات موضوع از قانون نزدیکی در گراف در بسیاری از معیارها همچون جاکارد، کسینوسی، شباهت شبکه، نرمال L1 و غیره استفاده شده است. در همه آن‌ها یک فرض ساده‌سازی به

طور مستقیم استفاده شده است، و آن هم این است که در این معیارها فقط دوست مشترک به طور مستقیم لحاظ می‌شود و فاصله‌های دورتر مانند دوست دوست مشترک، برای دو فرد لحاظ نمی‌شوند. علت این هست که در واقعیت تحقیقات نشان داده است که تاثیر دوست مشترک در جذب دو فرد به یکدیگر بسیار بالاتر از تاثیر دوست‌های مشترک، با واسطه‌های بیشتر می‌باشد.

فرض کنید در لحظه t در یک شبکه دو فردی که با یکدیگر ارتباط ندارند، دارای یک یا چند دوست مشترک می‌باشند. در لحظه t+1 در همان شبکه این دو فرد با یکدیگر دوست می‌شوند. دوست شدن دو فرد می‌تواند هم به شباهت آن دو و هم به نزدیکی آن دو در گراف دوستی وابسته باشد، بنابراین این دو مولفه مستقل از هم نمی‌باشند. اما با فرض ساده‌سازی استقلال آن دو از هم می‌توان ایجاد یال جدید را به نزدیکی در گراف و تاثیر دوست مشترک آن‌ها نسبت داد. این استدلال مشابه استدلال نیومن در به دست آوردن میزان تاثیر وزن هوموفیلی در شبکه که ایجاد یال بین افرادی با ویژگی‌های مشابه را مستقل از تاثیر نزدیکی در گراف دوستی آن‌ها در نظر گرفته‌اند، می‌باشد [۲۹].

با استدلال فوق می‌توان استنتاج کرد که به ازای هر سه تایی بسته در شبکه، یک یال جدید در شبکه تحت تاثیر دوست مشترک ایجاد شده است. بنابراین مشابه کار نیومن [۲۹] برای به دست آوردن میزان اهمیت دوست مشترک و نزدیکی در گراف کافیست تعداد سه تایی‌های بسته در شبکه را در قیاس با بیشینه تعداد حالات ممکن در شبکه، شمارش شود. برای این امر مطالعاتی صورت گرفت و مشخص شد برای این مسئله در ادبیات موضوع سه راهکار به شرح ذیل وجود دارد:

میانگین ضریب خوشه‌بندی محلی[1]: ضریب خوشه‌بندی محلی[2] برای هر گره تعریف شده و کیفیت اتصالات گره مذکور با همسایه‌هایش در قیاس با یک گره در گراف کامل[3] را تعیین می‌کند و از رابطه (۲) برای گره مشخص i حاصل می‌شود.

$$C_i = \frac{2e_i}{k_i(k_i-1)} \tag{۲}$$

که در آن $e_i$ برابر تعداد اتصالاتی که ما بین همسایه‌های گره مذکور i وجود دارد می‌باشد، همچنین $k_i$ برابر درجه گره i می‌باشد.

فرض کنید $e_i$ در شبکه‌ای برابر عدد k باشد، بنابراین k یال در شبکه ایجاد شده است که دو سر تمامی آن یال‌ها را همسایه‌های گره i تشکیل داده‌اند. درنتیجه تمامی گره‌های دو سر K یال، دارای حداقل یک دوست مشترک می‌باشند، که آن هم گره i می‌باشد. بنابراین براساس فرضیات فوق تمام k یال به خاطر نزدیکی در گراف و به خاطر واسطه‌گری گره i در شبکه ایجاد شده‌اند. لذا کافیست برای محاسبه وزن ویژگی ساختاری برای هر گره این مقدار را تقسیم بر مقدار بیشینه آن کنیم، که دقیقا برابر $C_i$ می‌شود. حال برای به دست آوردن وزن ساختاری کل شبکه کافیست میانگین $C_i$‌ها گرفته شود که برابر با میانگین ضریب خوشه‌بندی محلی می‌شود. رابطه (۳) نحوه محاسبه ضریب خوشه‌بندی محلی را نشان می‌دهد.

$$\bar{C} = \frac{1}{n}\sum_{i}^{n}\frac{2e_i}{k_i(k_i-1)} \tag{۳}$$

---

[1] Average Local Cluster Coefficient
[2] Local Clustring Cofficient
[3] Clique (Complete graph)

ضریب خوشه‌بندی سراسری[1]: معیاری است که میزان اتصالات کل گره‌های گراف را نسبت به گراف کامل متناظر آن می‌دهد. ضریب خوشه‌بندی سراسری از رابطه (۴) حاصل می‌شود.

$$C = \frac{\text{تعداد مثلث‌ها در گراف} \times 3}{\text{تعداد اتصالات سه تایی در گراف}} \quad (4)$$

موتیف سه تایی بسته[2]: در علوم شبکه مطالعاتی در رابطه با تعداد همبندی‌های m تایی در گراف با عنوان موتیف انجام می‌شود که در آن تعداد همبندی m تایی در شبکه با تعداد همبندی همان شبکه در حالت تصادفی و با حفظ توزیع درجه، قیاس می‌شود. در واقع موتیف به دنبال پیدا کردن فرکانس الگوهای ساختاری شبکه نسبت به حالت تصادفی می‌باشد. فرمول محاسبه موتیف، در رابطه (۵) آورده شده است. موتیف را به صورت $Motif_M(e)$ نشان می‌دهند، که در آن M تعداد گره‌های همبندی مورد مطالعه و e تعداد یال‌های همبندی مورد مطالعه می‌باشد. $n_M(e)$ تعداد M تایی‌هایی که e یال دارند را در گراف اصلی می‌شمارد. همچنین گراف تصادفی متناظر با گراف اصلی با حفظ درجه به صورت تصادفی چندین مرتبه ساخته می‌شود و میانگین و انحراف معیار همبندی‌های M تایی که e یال دارند، در آن شمرده می‌شود که در رابطه (۵) به ترتیب با $< n_M(e)^{Random} >$ و $\sigma_M^{Random}(e)$ نشان داده شده است [۳۰].

$$Motif_M(e) = \frac{n_M(e) - <n_M(e)^{Random}>}{\sigma_M^{Random}(e)} \quad (5)$$

از آنجا که میانگین ضریب خوشه‌بندی محلی به ازای هر دوست مشترک، تعداد یال‌های ایجاد شده تحت تاثیر آن را نسبت به بیشینه حالت ممکن شمارش می‌کند، لذا نزدیک‌ترین روش از سه روش فوق، این روش می‌باشد. در نتیجه نهایتا فرمول وزن ویژگی ساختاری به صورت رابطه (۳) محاسبه می‌شود.

بنابراین با توجه به شکل ۱ محاسبه وزن معیار هموفیلی با در اختیار داشتن اطلاعات کاربران توسط رابطه (۱) و محاسبه وزن معیار ساختاری با در اختیار داشتن تصویر لحظه‌ای شبکه در زمان t توسط رابطه (۳) به دست می‌آید. بنابراین فرمول نهایی تجمیع وزن‌دار در رابطه (۶) در ذیل ارائه شده است.

$$P(X,Y) = \begin{cases} \frac{\frac{W(c)}{W_{max}(c)} + (W_{Structural} \times NS(u,x))}{\frac{W(c)}{W_{max}(c)} + W_{Structural}} & , if\ X_k = Y_k \\ \frac{\left(\frac{W(c)}{W_{max}(c)} \times S_k(X_k,Y_k)\right) + (W_{Structural} \times NS(u,x))}{\frac{W(c)}{W_{max}(c)} + W_{Structural}} & , Otherwise \end{cases} \quad (6)$$

در این رابطه $W_{Structural}$ برابر رابطه (۴)، $NS(u,x)$ برابر معیار شباهت شبکه و $S_k(X_k,Y_k)$ برابر معیار شباهت OF می‌باشد. $X_k$ نیز مقدار ویژگی هموفیلی X برای گره X می‌باشد. همچنین $P(X,Y)$ برابر احتمال ایجاد یال بین دو گره X و Y می‌باشد.

## ۴- نتایج

در این بخش ابتدا در رابطه با نحوه جمع‌آوری و تکمیل دو مجموعه داده مورد استفاده در این تحقیق، توضیحاتی داده شده است، در ادامه به تبیین سناریوی ارزیابی پرداخته شده است. در نهایت به تحلیل و ارزیابی نتایج حاصل از دو مجموعه داده پرداخته شده است.

---

[1] Global Clustering Coefficient

[2] Motif (3)

## ۴-۱- مجموعه داده

برای این پژوهش دو مجموعه داده در نظر گرفته شده است که در ادامه هر یک تبیین خواهند شد.

**شبکه اجتماعی مجازی دانشگاه تحصیلات تکمیلی زنجان**[1]: این شبکه اجتماعی مجازی، کار خود را از سال ۸۹ شروع کرده و تا سال ۹۳ در فضای مجازی در دسترس بوده است. حدود ۶۰۰ دانشجو در دانشگاه تحصیلات تکمیلی زنجان مشغول به تحصیل بوده‌اند که از این بین، حدود ۳۰۰ نفر در شبکه اجتماعی مجازی این دانشگاه، به عضویت این وبسایت در آمدند. تمامی اطلاعات این وبسایت (در طول سال‌های ۸۹ تا ۹۳) جمع‌آوری شده است که شامل ۳۰۱ کاربر (شامل دانشجویان، اساتید و کارمندان) به همراه مشخصات وارد شده و همچنین ۳۱۸۲ ارتباطات مابین آن‌ها می‌باشد.

**تکمیل اطلاعات مشخصات کاربران**

بسیاری از مجموعه داده‌های جمع‌آوری شده از شبکه‌های اجتماعی مجازی به دلایل مختلف از قبیل پر نکردن تمام مقادیر پرسشنامه توسط کاربران در هنگام ثبت‌نام و یا فعال‌سازی سطح حریم خصوصی توسط کاربران (که در این صورت APIها قادر به استخراج اطلاعات اینگونه افراد نمی‌باشند) دارای نقصان اطلاعات می‌باشند. همین موضوع باعث شده است تا تکمیل اطلاعات ناقص در شبکه‌های اجتماعی مجازی اهمیت پیدا کند. از آنجا که در مجموعه داده جمع‌آوری شده نقصان مشخصات کاربران وجود داشت بنابراین ابتدا از طریق پرسشنامه مشخصات مربوطه کامل‌تر گردید. در ادامه برای تکمیل اطلاعات مشخصات کاربران از یکی از روش‌های تکمیل اطلاعات ناقص کاربران در شبکه‌های اجتماعی در [۱۲] بهره گرفته شد. در این روش با استفاه از قانون هوموفیلی اطلاعات ناقص تکمیل می‌شوند. روش کار به این صورت هست که اطلاعات ناقص ویژگی $i$ام (به طور مثال جنسیت) در حساب کاربری فرد $x$، به واسطه مقادیر ویژگی $i$ام در حساب کاربری دوستان مشترک آن فرد، تکمیل می‌شود. این کار با رای اکثریت انجام می‌شود. در واقع هر دوست فرد $x$، یک رای دارد. در نهایت برای رفع نقص ویژگی $i$ام در حساب کاربری فرد $x$ صفتی که بیشترین رای را کسب کرده است، انتخاب می‌شود. این درحالی هست که دو عامل محدود کننده برای پذیرش صفت وجود دارد:

1. کمترین تعداد رای‌ها ($f$): برای جلوگیری از نتیجه‌گیری به ازای تعداد آرای خیلی کم از یک عامل محدود کننده برای تعداد آرا استفاده شده است که با f نمایش داده شده است. در صورتی که تعداد رای‌های صفت برگزیده از یک حدی کمتر باشد، صفت پذیرفته نمی‌شود (ویژگی i برای گره x همچنان خالی باقی می‌ماند).

2. اکثریت قاطع در رای گیری ($t$): برای جلوگیری از نتیجه گیری به ازای درصدهای پایین شرکت کنندها، یک عامل محدود کننده برای درصد آرا استفاده شده است که با $t$ مشخص شده است. در صورتی که درصد آرا به ازای تعداد افراد شرکت کننده در رای گیری از یک حدی کمتر باشد آن صفت مورد قبول واقع نمی‌شود.

---

[1] http://coinlab.ut.ac.ir/resources

اگر به $f$ و $t$ مقادیری کمی داشته باشند، اطلاعات بیشتری تخمین زده می‌شود ولی دقت روش کاهش می‌یابد، همچنین اگر مقدار $f$ و $t$ زیاد باشد اطلاعات کمی تخمین زده می‌شود. بنابراین انتخاب مقادیر $f$ و $t$ حائز اهمیت می‌باشند. براساس رابطه (۷) بهینه‌ترین حالت برای این مقادیر در مجموعه داده آموزشی به دست می‌آید.

$$L(f,t) = Precision(f,t) \times Log(C(f,t)) \quad (۷)$$

که در آن $Precision(f,t)$ دقت تخمین به واسطه مقادیر $f$ و $t$ در مجموعه داده آموزشی می‌باشد و همین طور $C(f,t)$ تعداد تخمین‌های درست در مجموعه داده آموزشی می‌باشد. با مقداردهی متفاوت تابع فوق و ارزیابی بهترین نتیجه به صورت تجربی مقادیر بهینه برای $f$ و $t$ محاسبه می‌شوند. مقادیر $f$ و $t$ برای بیشینه $L$ بهینه‌ترین مقادیر ممکن می‌باشند. نتایج حاصل از شبیه‌سازی این روش تکمیل اطلاعات برای مجموعه داده شبکه اجتماعی دانشگاه تحصیلات تکمیلی زنجان در جدول ۳ قابل مشاهده می‌باشد. ویژگی جنسیت به علت نداشتن نقص و کامل بودن مقادیر آن برای تمام افراد در این جدول وارد نشده است. دقت هر ویژگی به ازای بهترین مقادیری که برای $f$ و $t$ محاسبه شده است به دست آمده است، همان‌طور که در این جدول ملاحظه می‌شود به علت پایین تکمیل اطلاعات ویژگی شهرستان محل سکونت این ویژگی در نظر گرفته نشده و بر روی آن اجرا گرفته نشده است. ستون «تعداد مقادیر ناقص قبل از اجرا» در این جدول، نشان دهنده تعداد نقص هر ویژگی پس از تکمیل اطلاعات به واسطه پرسشنامه می‌باشد. ستون چهارم در این جدول، نشان دهنده دقت روش اجرایی برای تکمیل اطلاعات ناقص می‌باشد که همان‌طور که مشخص است، دقت قابل قبولی به ازای تمام ویژگی‌ها به استثنای ویژگی شهرستان محل سکونت به ازای این روش حاصل شده است. از آنجا که در این پژوهش به دنبال بالا بردن دقت روش نمونه‌برداری لینک خودار می‌باشیم، و اطلاعات ناقص بر روی نتایج نهایی تاثیرگذار می‌باشد، تنها دو ویژگی «جنسیت» و «وضعیت تاهل» که در نهایت به صورت کامل و بدون اطلاعات ناقص در شبکه حاصل شده‌اند، برای این پژوهش در نظر گرفته شده است.

**شبکه اجتماعی مجازی پوکک[1]:** پوکک محبوب‌ترین شبکه اجتماعی مجازی در کشور اسلواکی می‌باشد که حتی پس از آمدن فیس‌بوک، محبوبیت این وب‌سایت نزول پیدا نکرده است.

جدول ۳: نتایج تکمیل اطلاعات ناقص مجموعه داده شبکه اجتماعی مجازی دانشگاه تحصیلات تکمیلی زنجان
Table 3: Result of complete missing data on IASBS Social

| | $f$ | $t$ | تعداد مقادیر پیش‌بینی شده | دقت | تعداد مقادیر ناقص قبل از اجرا | درصد مقادیر ناقص باقی مانده بعد از اجرا |
|---|---|---|---|---|---|---|
| وضعیت تاهل | 2 | 0.1 | 166 | 0.95 | 167 | 0 |
| مقطع تحصیلی | 1 | 0.5 | 81 | 0.79 | 131 | 16.5 |
| سال ورود به دانشگاه | 3 | 0.1 | 94 | 0.64 | 164 | 23 |
| رشته تحصیلی | 1 | 0.5 | 89 | 0.89 | 156 | 22 |
| شهرستان محل سکونت | 1 | 0.3 | این ویژگی به خاطر دقت پایین درنظر گرفته نشد. | 0.45 | 125 | - |

---

[1] Pokec

مجموعه داده این شبکه اجتماعی مجازی در طول ۱۰ سال جمع‌آوری شده است که شامل یک میلیون و ششصد هزار کاربر می‌شود [۳۱]. این مجموعه داده بر روی وب‌سایت دانشگاه استنفورد[1] قرار داده شده است.

به علت حجیم بودن این مجموعه داده یک نمونه‌برداری جستجوی اول سطح با شروع از قدیمی‌ترین گره از این مجموعه داده انجام شد و ۴۷۲۴۱ گره از این مجموعه داده به همراه ۸۹۴۷۷۶ یال از آن در این نمونه‌برداری، برداشت شد (برداشت با قدیمی‌ترین گره آغاز شده است بدین ترتیب می‌توان فرض کرد که شبکه‌ی حاصل از نمونه‌برداری، شبکه اجتماعی مجازی پوکک در سال‌های اولیه می‌باشد).

این مجموعه داده دارای ویژگی‌های زیادی می‌باشد ولی مقادیر اکثر آن‌ها بیش از پنجاه درصد دارای نقصان می‌باشند و تنها ویژگی که تمامی مقادیر برای آن وجود دارد، ویژگی جنسیت می‌باشد. بنابراین از آنجا که در این پژوهش اطلاعات ناقص بر روی نتایج نهایی تاثیرگذار می‌باشد تنها همین ویژگی از این مجموعه داده در نظر گرفته شده است.

## ۴-۲- سناریو ارزیابی

اصلی‌ترین معیارهای ارزیابی استفاده شده در حوزه پیش‌بینی یال عبارتند از: دقت[2]، فراخوانی[3]، معیار F[4]، صحت کلاس‌بندی[5]، مساحت زیر نمودار[6].

معیار صحت کلاس‌بندی به مسئله این پژوهش مربوط نمی‌شود، معیارهای دقت و فراخوانی و به طبع آن معیار F به علت اینکه برای محاسبه، همه‌ی یال‌های مشاهده نشده را بررسی می‌کنند، از نظر پیچیدگی زمانی، پیچیدگی بسیار بالایی دارند (هر سه معیار پیاده‌سازی شدند ولی در عمل به علت زمان‌بر بودن از استفاده از این معیارها صرف نظر شد که برای مجموعه داده‌های بزرگ با چالش همراه خواهند شد. بنابراین از بین معیارهای فوق، معیار مساحت زیر نمودار برای ارزیابی روش پیشنهادی انتخاب شد. از این معیار در ارزیابی بسیاری از کارهای علمی استفاده شده است [۱۴، ۳۲، ۳۳].

سناریو ارزیابی به این شرح است:

۱. الگوریتم زیر به تعداد ۱۰ مرتبه انجام می‌شود.

 a. تمام یال‌هایی که در گراف ایجاد نشده‌اند، در گروه یال‌های ایجاد نشده قرار داده می‌شوند.

 b. ده درصد از یال‌های شبکه به صورت تصادفی حذف می‌شوند و در گروه یال‌های مشاهده نشده قرار داده می‌شوند.

 c. وزن ویژگی‌های ساختاری و هموفیلی برای ۹۰ درصد باقی مانده گراف، محاسبه می‌شوند.

 d. به تعداد n بار روال زیر اجرا می‌شود:

---

i. به صورت تصادفی و مستقل یک یال از مجموعه یال‌های ایجاد نشده و یک یال از مجموعه یال‌های مشاهده نشده انتخاب می‌شود.

ii. براساس وزن‌های حاصل شده، احتمال یال منتخب از مجموعه یال‌های ایجاد نشده و مشاهده نشده براساس تجمیع وزن‌دار معیارهای شباهت محاسبه می‌شوند.

iii. اگر احتمال به دست آمده برای یال مشاهده نشده بیشتر از احتمال به دست آمده برای یال ایجاد نشده بود آنگاه:

$$n' = n' + 1$$

iv. در غیر اینصورت اگر احتمال‌های بدست آمده با هم برابر بودند آنگاه:

$$n'' = n'' + 1$$

v. با داشتن n و n' و n" مقدار AUC محاسبه می‌شود.

$$AUC = \frac{n' + 0.5n''}{n}$$

e. مقدار مجموع AUCها برای اجراهای متفاوت محاسبه می‌شود:

$$AUC_{Total} = AUC_{Total} + AUC$$

2. از ۱۰ اجرای گرفته شده میانگین گرفته می‌شود و مقدار نهایی AUC حاصل می‌شود:

$$AUC_{Finall} = \frac{AUC_{Total}}{10}$$

برای مجموعه داده شبکه اجتماعی مجازی دانشگاه تحصیلات تکمیلی زنجان در قسمت b از الگوریتم فوق به علت داشتن زمان تشکیل یال‌ها به جای حذف تصادفی یال‌ها، ده درصد از آخرین یال‌های تشکیل شده حذف شده‌اند، همچنین برای هر دو مجموعه داده n برابر نصف تعداد یال‌های حذف شده در نظر گرفته شده است (در واقع n برابر است با پنج درصد از کل یال‌های گراف اصلی).

## ۴-۳- نتایج و تحلیل داده

در ابتدا برای انتخاب معیار شباهت اصلح از میان معیارهای شباهت کاندید، مطالعه‌ای انجام دادیم که نتایج آن در ذیل آورده شده است.

### ۴-۳-۱- انتخاب معیار اصلح از میان معیارهای کاندید شباهت

نتایج بر روی مجموعه داده شبکه اجتماعی مجازی دانشگاه تحصیلات تکمیلی زنجان برای انتخاب معیار شباهت هوموفیلی برگزیده از بین دو معیار کاندید OF و IOF در جدول ۴ و برای انتخاب معیار شباهت ساختاری برگزیده از بین چهار معیار کاندید شباهت شبکه، کسینوس، نرمال L1 و جاکارد در جدول ۵ آورده شده است. همچنین نتایج مشابه بر روی مجموعه داده شبکه اجتماعی مجازی پوکک به ترتیب در جداول ۶ و ۷ آورده شده است.

**جدول ۴: ارزیابی معیار اصلح از بین معیارهای کاندید شباهت هوموفیلی برای مجموعه داده شبکه اجتماعی مجازی دانشگاه تحصیلات تکمیلی علوم پایه زنجان.**
**Table 4: The results of the evaluation of best similarity metrics between homophily similarity metrics on IASBS Social**

| معیار شباهت ساختاری | صفت مورد مطالعه | میانگین AUC های به دست آمده |
|---|---|---|
| OF | جنسیت | **0.55** |
|  | وضعیت تاهل | **0.52** |
| IOF | جنسیت | **0.53** |
|  | وضعیت تاهل | **0.52** |

برای به دست آوردن نتایج از معیار ارزیابی AUC که در بخش قبل توضیح داده شد، استفاده شده است. با این تفاوت که به جای استفاده از معیار شباهت تجمیع شده در آن از معیارهای مذکور استفاده شده است. همان‌طور که مطرح شد به علت تصادفی بودن الگوریتم به تعداد ۱۰ اجرا به ازای هر یک از معیارهای شباهت معیار AUC محاسبه شده، و در نهایت از آن میانگین گرفته شده است.

با توجه به نتایج حاصل از ارزیابی که در جداول ۴ و ۶ ارائه شده است، معیار شباهت OF توانسته است دقت بالاتری نسبت به معیار IOF کسب کند، بنابراین در این کار پژوهشی از بین معیارهای شباهت هموفیلی این معیار انتخاب شده است.

**جدول ۵: ارزیابی معیار اصلح از بین معیارهای کاندید شباهت ساختاری برای مجموعه داده شبکه اجتماعی مجازی دانشگاه تحصیلات تکمیلی علوم پایه زنجان.**
**Table 5: The evaluation of best similarity metrics between network similarity metrics on IASBS Social**

| معیار شباهت ساختاری | میانگین AUCهای به دست آمده |
|---|---|
| شباهت شبکه | 0.56 |
| جاکارد | 0.56 |
| شباهت کوسینوسی | 0.53 |
| نرمال L1 | 0.51 |

**جدول ۶: ارزیابی معیار اصلح از بین معیارهای کاندید شباهت هموفیلی برای مجموعه داده شبکه اجتماعی مجازی پوکک.**
**Table 6: The evaluation of best similarity metrics between network similarity metrics on IASBS Social**

| معیار شباهت ساختاری | صفت مورد مطالعه | میانگین AUCهای به دست آمده |
|---|---|---|
| OF | جنسیت | 0.49 |
| IOF | جنسیت | 0.46 |

همان‌طور که در جداول ۵ و ۷ نشان داده شده است، دو معیار شباهت شبکه و جاکارد، توانستند بیشترین دقت را نسبت به سایر معیارها در هر دو مجموعه داده کسب کنند (قابل ذکر است که اعداد نوشته شده همگی به دو رقم اعشار گرد شده‌اند). با توجه به برابر شدن دو معیار شباهت شبکه و جاکارد، تصمیم بر آن شد تا در این پژوهش از معیار شباهت شبکه برای تعیین شباهت ساختاری، بهره گرفته شود.

**جدول ۷: ارزیابی معیار اصلح از بین معیارهای کاندید شباهت ساختاری برای مجموعه داده شبکه اجتماعی مجازی پوکک.**
**Table 7: The evaluation of best similarity metrics between homophily similarity metrics on Pokec Social**

| معیار شباهت ساختاری | میانگین AUCهای به دست آمده |
|---|---|
| شباهت شبکه | 0.79 |
| جاکارد | 0.79 |
| شباهت کوسینوسی | 0.78 |
| نرمال L1 | 0.78 |

## ۲-۴-۳-۲- ارزیابی و تحلیل نتایج

در این بخش توسط معیار ارزیابی مطرح شده، به ارزیابی و تحلیل نتایج می‌پردازیم. ویژگی‌های شباهت به دو دسته هموفیلی و ساختاری تقسیم می‌شوند. دسته ویژگی ساختاری تنها شامل یک ویژگی می‌باشد، اما در دسته ویژگی‌های هموفیلی، تعداد ویژگی‌ها برابر با تعداد مشخصاتی است که از کاربران شبکه اجتماعی در دسترس می‌باشد. تخصیص وزن به معیارهای شباهت و تجمیع آن‌ها را از دو حیث می‌توان مطالعه نمود که در ذیل به آن اشاره می‌شود:

۱- زمانی که چندین ویژگی کاربران یک شبکه اجتماعی دردسترس باشد، برای استفاده از معیار شباهت هموفیلی باید نتایج هر یک محاسبه و با یکدیگر تجمیع شوند. چالشی که در اینجا مطرح می‌شود این است که آیا اهمیت هر یک از ویژگی‌ها در

انتخاب دوست تاثیر یکسانی دارد؟ در بسیاری از مطالعات صورت گرفته وزن هر یک از ویژگی‌های هموفیلی را یکسان در نظر گرفته‌اند. این امر در صورتی است که مطالعات اخیر نشان داده است این امر مبتنی بر واقعیت نمی‌باشد [34-36]. بنابراین می‌توان به واسطه وزن‌های به دست آمده برای ویژگی‌های هموفیلی به تجمیع وزن‌دار آن‌ها پرداخت و نتایج حاصل آن را با نتایج میانگین‌گیری ساده مقایسه کرد. نتایج حاصل به واسطه معیار ارزیابی مطرح شده بر روی مجموعه داده شبکه اجتماعی مجازی دانشگاه علوم پایه زنجان در جداول 8 و 9 ارائه شده است (با توجه به در دسترس بودن تنها یک ویژگی هموفیلی برای مجموعه داده شبکه اجتماعی مجازی پوکک، از این مجموعه داده در این بخش استفاده نشده است).

**جدول 8: نتایج ارزیابی به ازای هر ویژگی شباهت برای مجموعه داده شبکه اجتماعی مجازی دانشگاه تحصیلات تکمیلی علوم پایه**

Table 8: Evaluation results for each similarity feature on IASBS Social

| | میانگین AUCها | میانگین وزن‌ها |
|---|---|---|
| شباهت ساختاری (NS) | **0.56** | 0.48 |
| شباهت هموفیلی (جنسیت) | **0.55** | 0.29 |
| شباهت هموفیلی (وضعیت تاهل) | **0.52** | 0.24 |

همان‌طور که در جدول 9 مشاهده می‌شود، نتایج حاصل بر روی شبکه اجتماعی مجازی دانشگاه تحصیلات تکمیلی زنجان حاکی از آن است که با تخصیص وزن‌ها به ویژگی‌های جنسیت و وضعیت تاهل و استفاده از آن در تجمیع آن‌ها برای به دست آوردن معیار شباهت، توانسته‌ایم بر دقت این روش نسبت به زمانی که، وزن هر معیار مشابه در نظر گرفته شده است، 0/02 بیفزاییم.

از آنجا که وزن‌های حاصل شده اختلاف ناچیزی با یکدیگر دارند، در نظر گرفتن وزن‌ها در تجمیع در قیاس با حالت پایه (در نظر گرفتن وزن یکسان برای هر دو معیار) بهبود اندکی مشاهده شده است. طبیعتا هر چه اختلاف وزن‌ها برای معیارهای شباهت هموفیلی بیشتر باشند اختلاف دقت نتیجه روش ارائه شده با روش پایه بیشتر خواهد بود. همچنین از آنجا که وزن‌های تخصیص داده شده به ترتیب به شباهت ساختاری، شباهت هموفیلی برای ویژگی جنسیت و شباهت هموفیلی برای ویژگی وضعیت تاهل، بیشترین اعداد را نسبت داده است، انتظار داشتیم که معیار ارزیابی نیز بیشترین دقت را ابتدا به ویژگی ساختاری سپس به ویژگی هموفیلی جنسیتی و در نهایت به ویژگی هموفیلی وضعیت تاهل نسبت دهد، که نتایج منطبق بر همین امر می‌باشد.

**جدول 9: نتایج ارزیابی تجمیع ویژگی‌های هموفیلی برای مجموعه داده شبکه اجتماعی مجازی دانشگاه تحصیلات تکمیلی علوم پایه.**

Table 9: Evaluation results for aggregation homophily features on IASBS Social.

| | میانگین AUCهای به دست آمده |
|---|---|
| تجمیع ساده دو معیار هموفیلی | **0.55** |
| تجمیع وزن‌دار دو معیار هموفیلی | **0.57** |

2- محاسبه وزن‌های ویژگی‌های هموفیلی و ساختاری

علاوه بر حل چالش فوق (که توضیح داده شد)، می‌تواند در تجمیع ویژگی‌های هموفیلی و ساختاری کمک نماید، که این امر از اهمیت بیشتری برخوردار بوده و هدف اصلی این پژوهش می‌باشد. راهکار پیشنهادی به این شرح می‌باشد که ابتدا هر یک از ویژگی‌های هموفیلی و ساختاری با معیار شباهت متناظر خود محاسبه شوند و سپس به واسطه وزن‌های حاصل شده به ازای هر یک از آن‌ها با یکدیگر تجمیع شوند. همان‌طور که عنوان شد در

ادبیـات موضـوع برای مسـئله پیش‌بینی یال دو رویکرد وجود دارد (نزدیکی درگراف و هموفیلی)، با اسـتفاده از راهکار فوق می‌توان از مزایای هر دو رویکرد متناظر هر یک بهره‌مند شد.

### جدول ۱۰: نتایج تجمیع معیارهای هموفیلی و ساختاری برای مجموعه داده شبکه اجتماعی مجازی دانشگاه تحصیلات تکمیلی علوم پایه.
**Table 10: Evaluation results for aggregation homophily features and network features on IASBS Social.**

|  | میانگین AUCها |
|---|---|
| تجمیع ساده (جنسیت، وضعیت تاهل و ساختاری) | **0.57** |
| تجمیع وزن‌دار (جنسیت، وضعیت تاهل و ساختاری) | **0.60** |

در جدول ۱۰ نتایج تجمیع معیارها به صـورت وزن‌دار (راهکار پیشـنهادی) و میانگین‌گیری سـاده برای شبکه زنجان، ارائه شـده اسـت. روش پیشـنهادی توانسـته اسـت نسـبت به سـایر نتایج (معیار شـباهت جنسیت به تنهایی، معیار شباهت وضعیت تاهل به تنهایی، تجمیع سـاده دو معیار مذکور و تجمیع دو معیار مذکور به صـورت وزن‌دار) دقت را ۰/۰۳ افزایش دهد. همان‌طور که عنوان شـد نزدیکی وزن‌های حاصـله باعث شـده اسـت که میزان افزایش دقت قابل ملاحظه نشـود. در واقع میزان افزایش دقت مسـئله وابسـته به میزان تفاوت وزن ویژگی‌های حاصل شده (میزان تفاوت اهمیت ویژگی‌های شبکه) می‌باشد.

همان‌طور که در جدول ۱۱ مشـاهده می‌شـود وزن ویژگی هموفیلی جنسیت برای مجموعه داده شـبکه اجتماعی مجازی پوکک منفی شده است و همان‌طور که توضیح داده شد وزن منفی نشـان دهنده وجود هتروفیلی برای این ویژگی در این شـبکه می‌باشـد. در واقع در این شـبکه افراد تمایل بیشـتری به برقراری ارتباط با جنس مخالف از خود نشـان داده‌اند. از آنجا که معیار ارزیابی پیاده‌سازی شـده برای حالت تصادفی برابر است با مقدار ۵۰٪، لذا انتظار می رود که (با توجـه به اینکـه در این شـبکـه هتروفیلی وجـود دارد) برای معیار شـباهت هموفیلی جنسـیت این مقدار زیر ۵۰٪ حاصل شود.

### جدول ۱۱: نتایج وزن‌های حاصل شده به ازای هر ویژگی شباهت برای مجموعه داده پوکک
**Table 11: Results of the weights obtained per similarity feature on Pokec Social**

| | میانگین وزن‌ها | میانگین AUCها |
|---|---|---|
| شباهت ساختاری | 0.072 | **0.79** |
| شباهت هموفیلی (جنسیت) | -0.039 | **0.49** |

همان‌طور که در جدول ۱۱ مشاهده می‌شود مقدار مساحت زیر نمودار برای ویژگی هموفیلی جنسیت زیر ۵۰٪ حاصـل شـده اسـت که مبتنی با تحلیل نظری ارائه شـده می‌باشـد. از آنجا که معیارهای هموفیلی جهت اندازه‌گیری هموفیلی ارائه شـده‌اند و کارایی برای هتروفیلی نداشـته، اضـافه کردن ویژگی هموفیلی به سـاختاری طبیعتا باید دقت را کاهش دهد. همان‌طور که در جدول ۱۱ مشاهده می‌شود میزان مساحت زیر نمودار محاسبه شده برای ویژگی سـاختاری به تنهایی برابر ۰/۷۹ می‌باشـد که با اضـافه شدن معیار هموفیلی به صـورت وزن‌دار این عدد به ۰/۶۸ کاهش می یابد. با این حال همان‌طور که در جدول ۱۲ مشـاهده می‌شـود اگر اضـافه کردن معیار هموفیلی جنسـیت به صـورت میانگیری سـاده اضـافه شـود مقدار ۰/۷۹ به ۰/۶۵ کاهش می یابد. این امر بدان معناست که با احتسـاب اینکه در شـبکه هتروفیلی وجود دارد ولی میزان اهمیت آن از ویژگی سـاختاری کمتر بوده (کمتر بودن وزن ویژگی هموفیلی جنسـیت از ویژگی سـاختاری در جدول ۱۱) بنـابراین در تجمیع وزن‌دار از میزان اهمیت آن نسـبت به

میانگین‌گیری ساده کاسته شده و توانسته است نزول کمتری پیدا کند.

**جدول ۱۲: نتایج تجمیع معیارهای هموفیلی و ساختاری برای مجموعه داده پوکک**
**Table 12: Evaluation results for aggregation homophily features and network features on Pokec Social.**

| میانگین AUCها | |
|---|---|
| 0.65 | تجمیع ساده (معیار هموفیلی و ساختاری) |
| 0.68 | تجمیع وزن‌دار (معیار هموفیلی و ساختاری) |

## ۵- نتیجه‌گیری

در این پژوهش در ابتدا مجموعه داده شبکه اجتماعی مجازی دانشگاه تحصیلات تکمیلی زنجان را جمع‌آوری کردیم. این مجموعه داده دارای نقص اطلاعات کاربران بود. برای حل آن از طریق پرسشنامه و در نهایت به واسطه پیاده‌سازی یکی از روش‌های تکمیل اطلاعات در حوزه علوم شبکه این نقص اطلاعات را برطرف نمودیم.

براساس مطالعات صورت گرفته، معیارهای شباهتی از هر دو دسته معیارهای شباهت هموفیلی و ساختاری به عنوان کاندید در نظر گرفتیم و با ارزیابی نتایج روی دو مجموعه داده، بهترین معیارها را برگزیدیم.

در این پژوهش توانستیم با ارائه یک مدل به تجمیع وزن‌دار معیارهای شباهت هموفیلی و ساختاری بپردازیم. همچنین به واسطه آن توانستیم به دقت بهترین نتایج معیارهای مبتنی بر همسایگی گره بیفزاییم. از مزایای این کار پژوهشی عدم نیاز به دسترسی به کل گراف در همان لحظه می‌باشد و فقط تصویر لحظه‌ای شبکه در زمان گذشته می‌تواند کفایت کند.

در شاخه علوم انسانی به علت تعدد زیاد ویژگی‌های هموفیلی انتخاب تعدادی از ویژگی‌های هموفیلی خود یکی از مسائل و چالش‌ها محسوب می‌شود. از نتایج این پژوهش می‌توان در حل این مشکل استفاده کرد. همچنین در حوزه تشخیص اجتماعات روش‌هایی وجود دارد که با استفاده از هموفیلی و نزدیکی در گراف به تشخیص اجتماعات می‌پردازد. با استفاده از نتایج این پژوهش و تخصیص وزن‌های هموفیلی و نزدیکی در گراف می‌توان به دقت این روش‌ها کمک کرد. در این پژوهش معیاری برای هتروفیلی معرفی نشد که می‌توان به عنوان کارهای آتی به آن پرداخت و نتایج این پژوهش را غنی‌تر نمود.

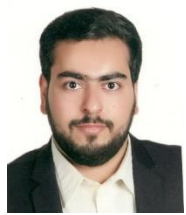

**علیرضـا اسـحاق پور** متولد ۱۳۶۹، کارشناسی و کارشناسی ارشد خود را به ترتیب در سال‌های ۱۳۹۲ و ۱۳۹۶ در دانشـگاه IASBS و دانشـگاه تهران به پایان رسانید. وی هم اکنون به عنوان کارشناس ارشد SOC با هلدینگ بهین راهکار همکاری می‌نماید. زمینه‌های پژوهشـی مورد علاقه ایشـان شـبکه‌های اجتماعی مجازی، الگوریتم‌های ژنتیک چند هدفه و امنیت اطلاعات می‌باشد.

نشانی رایانامه ایشان عبارت است از:

**a_eshaghpoor@ut.ac.ir**

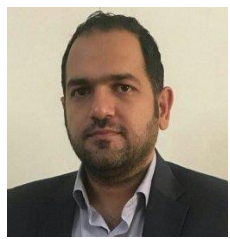

**مصـطفی صـالحی** کارشـناسـی و کارشـناسی ارشد خود را در رشته مهندسـی کامپیوتر به ترتیب در سـال‌های ۱۳۸۴ و ۱۳۸۶ به پایان رسـانید. او در سـال ۱۳۹۱ موفق به دریافت درجه دکتری در همین رشـته از دانشگاه شـریف شـد. پس از آن پسـا دکتری خود را به ترتیب در سـال‌های ۱۳۹۴ و ۱۳۹۵ در دانشـگاه بولونیا کشـور ایتالیا و دانشـگاه تلکام سـود پاریس (فرصـت مطالعاتی) گذراند. وی هم اکنون به عنوان عضـو هیئت علمی گروه بین رشـته‌ای فناوری دانشگاه تهران، با مرتبه دانشیاری مشـغول به فعالیت اسـت. زمینه‌های پژوهشـی ایشـان شـامل شـبکه‌های اجتماعی مجازی، اینترنت اشـیا و شـبکه‌های چندلایه می‌باشد.

نشانی رایانامه ایشان عبارت است از:

**mostafa_salehi@ut.ac.ir**

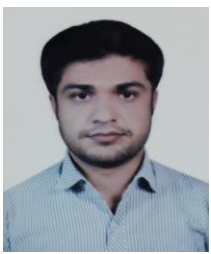

**وحیـد رنجبر** کـارشـــنـاســـی و کارشـناسی ارشد خود را در رشته مهندسـی فناوری اطلاعات به ترتیب در سـال‌های ۱۳۹۰ و ۱۳۹۲ به پایان رسانید. وی در سال ۱۳۹۷ دکتری تخصصی خود را در رشته فناوری اطلاعات در دانشـگاه تهران دریافت کرد. وی هم اکنون به عنوان عضـو هیئت علمی دانشـکده مهندسـی کامپیوتر دانشـگاه یزد مشـغول به فعالیت اسـت. زمینه‌های پژوهشـی مورد علاقه ایشان تحلیل شـبکه‌های اجتماعی ، یادگیری ماشین و کلان داده می‌باشد.

نشانی رایانامه ایشان عبارت است از:

**vranjbar@yazd.ac.ir**